# Experimentally Mapping the Phase Diagrams of Photoexcited Small Polarons

*Jocelyn L. Mendes and Scott K. Cushing**

Division of Chemistry and Chemical Engineering, California Institute of Technology, Pasadena, California 91125, United States

*Correspondence and requests for materials should be addressed to S.K.C. (email: scushing@caltech.edu)

ABSTRACT

Understanding the fundamental properties that dictate photoexcited polarons in materials is critical to tuning their properties. Theoretical models of polarons have only recently been extended to the excited state. Experimental measurements of polaron formation and transport have been widely undertaken across a range of materials, from photocatalysts and superconductors to soft conducting polymers. Here, we map thermalized excited state experimental measurements of quantities such as polaron strength onto phase diagrams of the Holstein, Hubbard–Holstein, and t-J–Holstein models. This work demonstrates that tuning electron–phonon coupling strength, electron localization, and spin exchange can be leveraged to suppress or control polarons in transition metal oxides. We find that the t-J–Holstein model provides the clearest qualitative description of electron polarons in the measured iron oxides and could be generally applied to a wide range of systems that exhibit polaron formation in the excited state. This work combines experimental data with ground state models to provide a qualitative parameter space for informing photoexcited electron polaron design, under which excited state polaronic behavior can be classified within ground-state calculable models.

MAIN TEXT

## Introduction

Polarons dominate material physics, ranging from solar energy materials,[1] O-LEDs,[2] organic semiconductors,[3] conducting polymers,[4] and even superconductors.[5] A polaron is the localization of a charge carrier due to coupling with local lattice deformations. The extent of the charge carrier's localization results in a distinction between small and large polarons, but this definition is often ambiguous.[6] It is generally accepted that a small polaron is localized to a unit cell or less in a strong coupling perturbation limit, while large polarons span multiple unit cells; however, the crossover between small and large polarons is an active area of study.[6,7] Extensive work has been undertaken both experimentally and computationally to understand and model polaronic behavior in emergent materials, ranging from scanning probe microscopy and core-level spectroscopies to numerical solutions of the Holstein model, which describes polaronic mobility.[8–12] Small polarons have been measured and theoretically described in the ground state for decades.[13–16] Here, we focus on understanding the fundamental properties of *photoexcited* polarons in solid-state materials, an area that remains less understood, despite a range of measurements of individual systems.

Photoexcited polarons form when a charge transfer transition redistributes the charge density between two local orbitals and the carriers couple to lattice vibrations. For example, in a transition metal oxide that is a charge transfer insulator, photoexcitation transfers an electron from a dispersive O 2p orbital to a localized transition metal d-orbital. Usually, during the photoexcited carrier thermalization time, the polarizable lattice of the transition-metal oxide locally distorts, trapping the charge carrier. This has been widely studied in the candidate photoelectrode hematite ($\alpha$-$Fe_2O_3$), showing that small-polaron formation occurs within <100 fs during the first electron–optical-phonon scattering event.[10,17,18] It has been further demonstrated that though these polarons form within the first electron–optical-phonon scattering event, the resulting localized polaronic state dominates carrier transport and persists up to carrier recombination and relaxation.[17] Polaron localization reduces carrier mobility and photoconversion efficiency, limiting the solar energy and fuel applications of hematite despite its otherwise favorable energetic properties.

Since initial measurements on $\alpha$-$Fe_2O_3$, a wide range of transition metal oxides have been measured to modify, understand, and suppress small polaron formation in strongly polar semiconductors.[17–21] Altering the nature of the local lattice, the metal center, surface functionalization, and spin correlations has reduced the strength and even suppressed small polarons in this class of materials. However, succinct design rules that can be used to predict new materials or extend ground state small polaron properties to the excited state have yet to be realized.

Here, we map a range of experimental photoexcited small polaron measurements in transition metal oxides onto three ground-state models: the Holstein, Hubbard–Holstein, and t-J–Holstein Hamiltonians. The resulting phase diagrams are constructed using ultrafast X-ray spectroscopy techniques that measure polaron and phonon energies which can be used to approximate the polaron binding energy and electron–phonon coupling strength, supplemented by experimentally derived exchange energies per spin in literature. We focus particularly on iron oxide systems to standardize metal centers for comparison of the results. Additionally, we consider only measurements of electron polarons on metal centers within these transition metal oxides, due largely to the fact that more X-ray spectroscopy-based measurements of electron polarons exist in literature. While hole polarons are likely also forming on ligand sites during the photoexcited polaron formation process in these materials, it is challenging to measure both simultaneously. We demonstrate that commonly employed scaling factors for polaron mobility and strength provide useful descriptors of the thermalized excited states considered here when constructed from the experimental and ground-state quantities described below. The Holstein model effectively maps long-lived and thermalized excited state dynamics onto the small-to-large polaron transition but provides limited insight beyond the role of the electron–phonon coupling strength. The Hubbard–Holstein model accounts for on-site repulsion ($U$) and well describes the small-to-large polaron transition, but it predicts bipolaron regimes that are rarely observed experimentally in the systems considered here. The t-J–Holstein model, related to the Hubbard–Holstein model in the strong coupling limit, provides the most insight by qualitatively describing when superexchange competes with electron–phonon coupling to reduce polaron localization in the excited state. These findings highlight that on-site electron-spin correlations are critical design parameters for controlling polarons in the photoexcited state. The resulting phase diagram parameterizations provide an applicable thermalized excited-state descriptor not only for solid-state systems but also molecular semiconductors,[22] and therefore we expect it to find wide use across solar energy materials,[23] soft conducting polymers,[4] and even insight into correlated quantum phases like superconductors.[24,25]

**Theory**

In the ground state, polarons at their simplest can be described by the Holstein Hamiltonian:[7,26]

$$H = -t \sum_{\langle ij \rangle \sigma} (c_{i\sigma}^{\dagger} c_{j\sigma} + c_{j\sigma}^{\dagger} c_{i\sigma}) + \omega_0 \sum_{i} b_i^{\dagger} b_i + g \sum_{i,\sigma} n_{i\sigma} \left( b_i + b_i^{\dagger} \right) \quad (1)$$

where $t$ is the hopping integral, $c_{i\sigma}^{\dagger}$and $c_{j\sigma}$ are the creation and annihilation operators, respectively, with spin, σ, for an electron at sites $i$ and $j$. $n_i$ is the number operator for electron spin, $g$ corresponds to the electron–phonon coupling strength, and $\omega_0$ refers to the phonon frequency. $b_i^{\dagger}$ and $b_i$ are the creation and annihilation operators, respectively, for phonons at site $i$.

The Holstein Hamiltonian, while insightful regarding the balance of electron–phonon coupling strengths, phonon frequency, and hopping integral, does not sufficiently capture the physics of complex transition-metal oxides. A Hubbard $U$ term can be added to the Holstein model to accurately describe strong on-site repulsions of the d orbitals, particularly at half-filling, as is the case for most of the Fe(III) oxides considered in this work.[27,28] The ratio of the charge hopping integral, $t$, to $U$ influences the orbital mixing within the band structure. It determines whether the material is a charge transfer insulator or a Mott insulator.[29] In a charge transfer insulator, the upper valence bands (VBs) are dominated by ligand orbitals, and metal d orbitals dominate the lower conduction bands (CBs). In a Mott insulator, the upper VBs and lower CBs are dominated by metal d orbitals due to the emergence of the Hubbard $d$ band. $U$ contributes to carrier localization and thus modulates polaron formation. A representative Hubbard–Holstein Hamiltonian can be written as:

$$H = -t\sum_{\langle ij\rangle\sigma}(c_{i\sigma}^{\dagger}c_{j\sigma} + c_{j\sigma}^{\dagger}c_{i\sigma}) + U\sum_{i} n_{i\uparrow}n_{i\downarrow} + \omega_0\sum_{i} b_i^{\dagger}b_i + g\sum_{i,\sigma} n_{i\sigma}\left(b_i + b_i^{\dagger}\right) \quad (2)$$

This Hamiltonian can be further downfolded into a t-J–Holstein model in the limit $U >> t$, which holds for the measured transition metal oxides in this work.[30,31] This downfold includes the spin-exchange integral ($J$) and, for antiferromagnetic (AFM) materials, describes superexchange in which a spin can undergo indirect exchange mediated by a ligand between two metal sites. AFM superexchange is favored when metal-ligand-metal bond angles are close to 180° and scales with the bond distance and bond angles, as described by Goodenough-Kanamori rules.[32–34] At half-filling, $J = 4t^2/U$. Thus, to a first order approximation, $4/(J/t) \approx U/t$ in a strong coupling limit. If $t + J >> g$, then no polaron forms, and photoexcited carriers thermalize to the band edge. If $t + J << g$, a polaron forms – often without significant prior thermalization. As $J$, $t$, and $U$ are all interrelated, modifying one through materials design will modify the others, and thus a balance must be struck between the parameters when considering materials design that aims to tune polarons. The downfolded t-J–Holstein Hamiltonian can be written as:

$$H = -t \sum_{\langle ij \rangle \sigma} (c_{i\sigma}^{\dagger} c_{j\sigma} + c_{j\sigma}^{\dagger} c_{i\sigma}) + J \sum_{ij} \vec{S_i} \cdot \vec{S_j} + \omega_0 \sum_{i} b_i^{\dagger} b_i + g \sum_{i,\sigma} n_{i\sigma} (b_i + b_i^{\dagger}) \qquad (3)$$

Each Hamiltonian outlined above describes different polaronic coupling regimes and models polaronic properties accordingly, though all are derived from a Holstein framework. The consideration of on-site repulsions and spin exchange in the Hubbard–Holstein and t-J–Holstein models, respectively, increases the complexity of each model Hamiltonian, but also acts to more accurately predict the polaronic properties of strongly electron- and spin-correlated materials.

**Methods: Polaron Parameterization from Experimental Measurements**

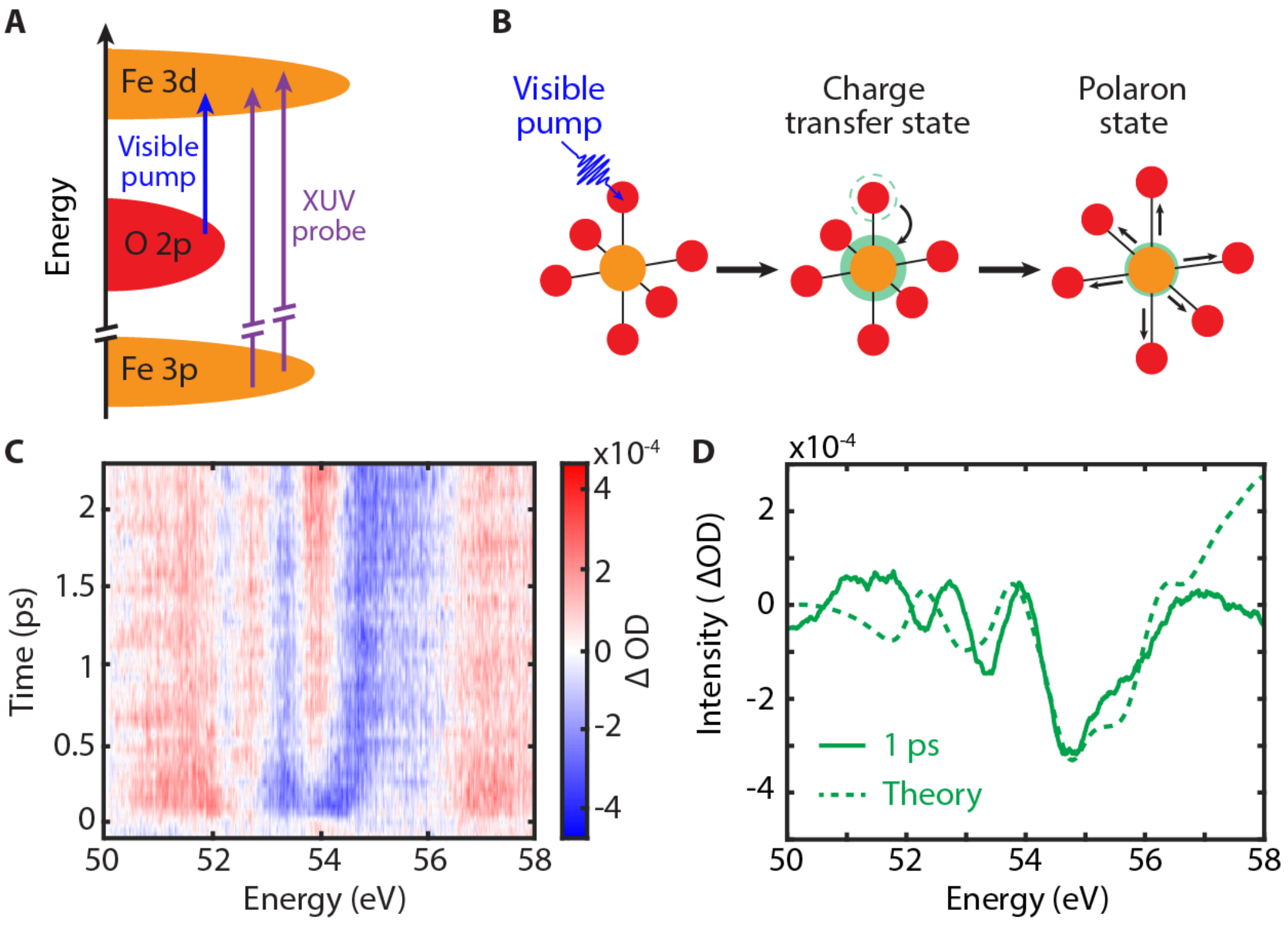

**Figure 1. Transient XUV spectroscopy probes the Fe $M_{2,3}$ edge to measure polaron formation. a.** Schematic representation of the 3p to 3d core-valence transition being probed during a transient XUV experiment following an above band gap photoexcitation. **b.** Diagram of the carrier localization and structural distortion caused by photoexcited small electron polaron formation where photoexcitation typically induces ligand-to-metal charge transfer (LMCT). The excited electron density (green) on the metal sites couples to a lattice mode that locally expands the $FeO_6$ octahedron and further localizes the carriers. **c.** Representative transient XUV reflection-absorption spectra at the Fe $M_{2,3}$ edge in $GdFeO_3$ centered at ~54 eV. At pump-probe overlap ($t_0$), a series of positive and negative peaks are observed. Immediately after photoexcitation with an 800 nm pump, an ultrafast spectral blueshift is observed around ~55 eV, which is attributed to octahedra expansion and small polaron formation. **d.** Representative experimental lineout after polaron formation is complete (1 ps) compared with DFT+BSE calculated polaronic spectra.

While these models, especially the Holstein model, have been widely applied to ground state polaronic properties, they are rarely compared to excited state polaron dynamics. Extensive nonequilibrium modeling of photoexcited polaron formation and trapping suggests that equilibrium models are insufficient for capturing the ultrafast dynamics of the first electron–phonon scattering event, during which photoexcited charge transfer and polaron formation occur.[1,11,35–37] Here, we consider parameterizing excited-state experiments within the framework of these ground-state model Hamiltonians, using thermalized and localized polaronic states. The spectral shifts associated with these binding energies persist for tens to hundreds of picoseconds and do not recover to their ground states within the temporal window of the experiments considered. Therefore, we consider these long-lived spectral shifts to be thermalized when considering their role in the photoexcited polaron formation and applicable to the ground-state Hamiltonian framework considered herein. Here, "thermalized" refers to equilibration within the electronic and local vibrational degrees of freedom that define the polaron state itself.

We employ transient extreme ultraviolet (XUV) reflection-absorption spectroscopy to measure the photoexcited polaron energy and phonon frequency. Transient XUV employs an above band gap pump to photoexcite samples and a core-valence probe that is sensitive to changes in electronic structure, local coordination, and oxidation state. An example of the experimental characterization is shown in Figure 1.

In the iron oxide systems considered here, a 3p to 3d core-to-valence transition is probed at the Fe $M_{2,3}$ edge. Figure 1A schematically represents the core-valence transition being probed following above band gap excitation. The resulting transient XUV spectrum is directly sensitive to the changes of the angular momentum coupling, electronic screening, and structural distortions following photoexcitation at the Fe $M_{2,3}$ edge. Upon photoexcitation, electron polarons are formed and localize carriers onto Fe sites and induce anisotropic lattice distortions, expanding Fe–O bond lengths (Figure 1B). This reduces the overlap between Fe 3d and O 2p orbitals, further localizing the electron polaron at the Fe site. As a result, a higher energy XUV photon is necessary to induce the Fe 3p to 3d core-valence transition following photoexcitation, which is why a transient blueshift is observed. The magnitude and kinetics of this shift report on the extent and rate of the photoexcited polaron localization (Figure 1C).

These spectral shifts dominate the transient spectra out to hundreds of picoseconds to nanoseconds, suggesting that the localized polaronic state dominates these spectra well after initial formation and hot carrier thermalization. The spectral shifts associated with photoexcited polaron formation dominate the spectra such that we can ascribe them to be quasi-equilibrated out to long timescales, as has been demonstrated in perovskite systems.[38,39] Further, pump-fluence-dependent studies of photoexcited polarons using transient XUV spectroscopy find that the transient dynamics are independent of fluence, though high excitation densities are typically required in these experiments to resolve signal.[40]

This spectral shift to higher energies has been verified through a range of theoretical predictions of X-ray edges, extending from semi-empirical charge transfer multiplet theory to first-principles density functional theory (DFT) and Bethe-Salpeter equation (BSE) theory.[41] In our work, we validate the differential spectrum of the polaron feature using a DFT+BSE calculated X-ray spectrum that accounts for photoexcited carriers and polaronic lattice distortions (Figure 1D).[20,41–45] The spectrum is then decomposed into the appropriate excited states to extract the related kinetics using singular value decomposition (Fig. 2). Heating effects are deconvolved from the spectra by modeling isotropic lattice expansions associated with lattice heating models and are not found to meaningfully contribute to the measured dynamics out to nanosecond timescales.[20]

We particularly expect electron polaron formation in charge transfer insulating iron oxides, which have conduction bands dominated by iron *d* orbitals. An above band gap excitation would thus induce a ligand-to-metal charge transfer (LMCT) from ligand dominated valence bands to metal dominated conduction bands, forming electron polarons on metal sites and hole polarons on ligand sites. Here, we cannot experimentally resolve the O K edge and therefore consider only electron polarons. This polaron formation may differ for intermediate and Mott insulators, which have different accessible excitation manifolds. These intermediate and Mott insulators may have metal-to-metal charge transfer (MMCT) transitions induced by photoexcitation, where electron and hole polarons can both be formed on metal sites.

As shown in Figure 2, a variety of polaron formation kinetics have been measured using transient XUV spectroscopy. In hematite, polaron formation occurs within the first electron–optical-phonon scattering time of ~100 fs.[17] In $CuFeO_2$, a polaron forms on the same timescale as hematite, but a *c*-axis lattice expansion on a picosecond timescale enables a delocalization of the initial polaron state.[42] In $ErFeO_3$, coherent charge hopping occurs before dephasing, and polarons are formed on a 2.3 ± 0.3 ps timescale, indicating large-polaron or mixed polaron/free electron conduction behavior.[44] $GdFeO_3$ does not form a polaron under 400 nm photoexcitation of a ligand-to-metal charge transfer (LMCT) that preserves the high-spin ground state. In contrast, 800 nm photoexcitation of a metal-to-metal charge transfer (MMCT) forms polarons in 250 ± 40 fs.[20]

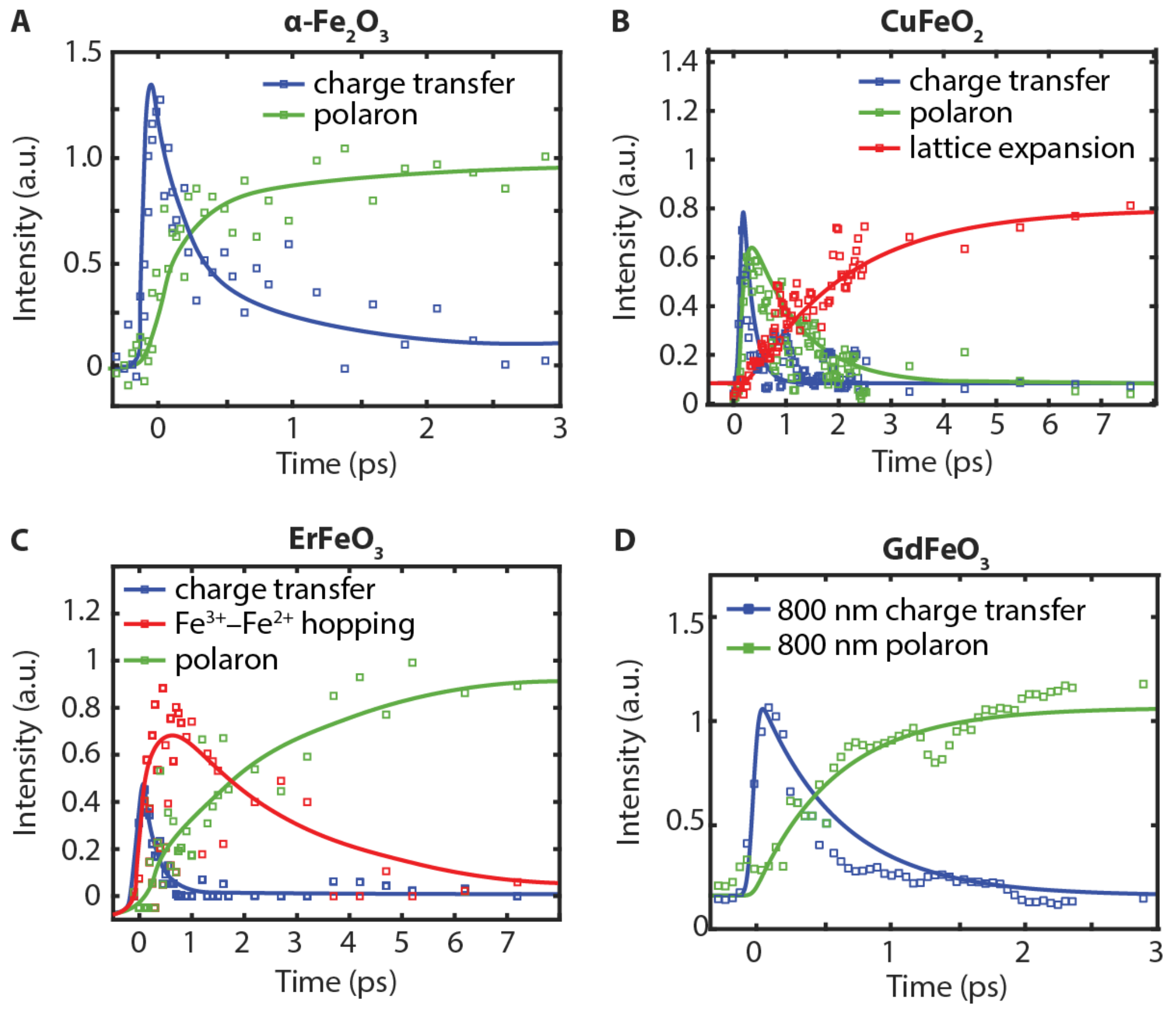

**Figure 2. Formation timescales and dynamics for polaron-forming Fe–O-based materials. a.** Hematite forms a polaron within ~100 fs, leaving few free carriers. **b.** $CuFeO_2$ forms a polaron within ~100 fs, which relaxes on a picosecond timescale through a *c*-axis expansion and subsequent delocalization. **c.** $ErFeO_3$ forms a polaron only after a period of $Fe^{2+}$–$Fe^{3+}$ charge hopping associated with superexchange. **d.** $GdFeO_3$ only forms a polaron following an 800 nm metal-to-metal charge transfer (MMCT) whereas no polarons form following a 400 nm ligand-to-metal charge transfer due to superexchange mechanisms under different spin configurations.

We have found, consistent with optical reflectivity experiments,[46] that the spectral blueshift of the XUV signal corresponding to polaron formation is approximately twice the polaron binding energy ($E_{polaron}$), to a first order approximation. These spectral shifts are long-lived and represent the thermalized polaron signal in the XUV spectra following formation, rather than the dynamics of the trapping process itself. $E_{polaron}$ is calculated from the magnitude of the spectral shift at the Fe $M_{2,3}$ edge, and $\omega_0$ is identified from the mode associated with the distortion of the lattice following polaron formation. For most materials, $\omega_0$ is on the order of the highest optical phonon frequency, which is fairly consistent for iron oxide-based materials at 80–100 meV.[47] The polaron energy can be described by Equation 4:

$$E_{polaron} = -\frac{g^2}{w_0} + t\exp\left(-\frac{g^2}{\omega_0^2}\right)cos(k) + \cdots \qquad (4)$$

where $k$ is momentum. For the materials measured here, except $ErFeO_3$ which exhibits large-polaron-like behavior, we can truncate the sum of the polaron energy in the first term due to the small electronic bandwidths of the measured iron oxides. We evaluated the next bandwidth-narrowing term, $t\ exp(-g^2/\omega_0^2)\ cos(k)$ and found that for the small-polaron-forming materials considered in this work, this correction is less than 1% of $E_{polaron}$, supporting the first-term truncation. $ErFeO_3$ remains a boundary case, with a larger correction of approximately 25% and represents an instance where different models may be necessary for large polaron approximations.

Using the parameterization of experimental XUV spectra and the truncation of Equation 4, we can estimate the electron–phonon coupling parameter, $g$. DFT+U calculations are used to approximate the electronic hopping integral $t$.[48,49] Note that the expression for electronic bandwidth goes as $2dt$ where $d$ is the dimensionality, here $d$ is taken as three. We employ a ground-state approximation of $t$ however we note that during photoexcitation there can be subtle changes that modify $t$, where intuitively it is found that the electron hopping decreases as polaron formation and renormalization occur, due to localization effects.[50] Direct experimental measurements of the renormalization of the electronic hopping integral remain scarce in literature, and excited-state calculations require computationally demanding, method-dependent quantum dynamics methods.[51–53] A Lang-Firsov approach can be used to estimate the effective electronic hopping integral ($t_{eff}$) following photoexcited polaron formation, however, because we approximate the electronic hopping integral ($t$) from ground-state theory, we consider the undressed hopping integral, rather than direct approximations of the polaronic excited state.[7,54,55]

The measured and calculated parameterization of the scalar quantities of the Holstein, Hubbard–Holstein, and t-J–Holstein models for a range of polaron forming transition metal oxides are given in Table 1, following the above outlined parameterization. The resulting values in Table 1 are conditional upon the assumptions discussed above, including the small-polaron truncation used to infer $g$ and the use of the ground-state bandwidth for $t$.

**Table 1.** Model parameters used in the Holstein, Hubbard–Holstein, and t-J–Holstein phase diagram mapping. All values are in units of eV. $J$ is approximated as exchange energy per spin. A detailed discussion of the errors associated with these parameters can be found in the Supporting Information. "–" denotes no known literature value for one of the necessary parameters.

| | $\lvert E_{polaron} \rvert$ | $\omega_0$ | $2dt$ | $J$ | $g = \sqrt{E_{polaron*w_0}}$ | $\lambda = \frac{g^2}{2dt\omega_0}$ | $\gamma = \frac{\omega_0}{2dt}$ | $g/2dt$ | $4/\left(\frac{J}{t}\right) = U/t$ |
|---|---|---|---|---|---|---|---|---|---|
| $\alpha$-$Fe_2O_3$ | 0.50 ± 0.05 [17] | 0.08 [56] | 0.19 [41] | 0.0097 ± 0.0031 [57,58] | 0.20 ± 0.01 | 2.63 ± 0.26 | 0.42 | 1.05 ± 0.05 | 13.07 ± 4.18 |
| $CuFeO_2$ | 0.35 ± 0.05 [42] | 0.09 [59] | 0.34 [42] | 0.025 ± 0.018 [60–62] | 0.18 ± 0.013 | 1.03 ± 0.15 | 0.26 | 0.52 ± 0.04 | 9.21 ± 6.63 |
| $ErFeO_3$ | 0.05 ± 0.005 [44] | 0.02 [44] | 0.91 [44] | 0.38 ± 0.16 [63] | 0.03 ± 0.002 | 0.05 ± 0.005 | 0.02 | 0.03 ± 0.0015 | 1.59 ± 0.67 |
| $BiFeO_3$ | 0.65 ± – [64] | 0.07 [64–66] | 0.33 [64] | 0.17 ± 0.05 [67–70] | 0.21 ± – | 1.97 ± – | 0.21 | 0.65 ± – | 1.32 ± 0.38 |
| $\alpha$-FeOOH | 0.55 ± – [71] | 0.08 [71–73] | 0.22 [74] | 0.19 ± – [75] | 0.21 ± – | 2.50 ± – | 0.36 | 0.95 ± – | 0.79 ± – |
| $GdFeO_3$ | 0.20 ± 0.08 [20] | 0.07 [47] | 0.13 [20] | 0.13 ± 0.04 [76,77] | 0.12 ± 0.02 | 1.54 ± 0.62 | 0.54 | 0.91 ± 0.18 | 0.68 ± 0.20 |
| a-$TiO_2$ | 0.10 ± – [78, 94] | 0.11 [79,80] | 1.51 [81] | – | 0.10 ± – | 0.07 ± – | 0.07 | 0.07 ± – | – |
| $BiVO_4$ | 0.44 ± – [82] | 0.10 [83] | 0.79 [84] | – | 0.21 ± – | 0.56 ± – | 0.13 | 0.27 ± – | – |

For some of the systems described in Table 1, it is important to consider superexchange and the Hubbard interaction. Of the above materials, $GdFeO_3$ is the only material that is not a charge transfer insulator. A charge transfer insulator has a valence band (VB) dominated by O 2p orbitals and a conduction band (CB) primarily composed of Fe 3d orbitals. $GdFeO_3$ is an intermediate insulator, meaning that the VB is composed of mixed Fe and O orbital character, the CB is dominated by Fe d orbitals much like charge transfer insulators. This enables selective excitation of both LMCT and MMCT transitions.

To determine the magnitude of the exchange integral, $J$ is taken as the exchange energy per spin, which is equivalent to $zJ_1S^2$ where $z$ is the connectivity, $J_1$ corresponds to values derived from inelastic neutron scattering (INS) and Raman scattering experiments in literature, and $S$ corresponds to the spin of the system.[85,86] Further details on the experimentally derived exchange energies and their associated uncertainties can be found in the Supporting Information. Superexchange at half-filling goes as $J = 4t^2/U,$ so changes to either $J$ or $U$ will modulate polaron formation. U/t is thus 4/(J/t) in the strong coupling limit of small polarons.[87,88] This relation is used in this work as a qualitative mapping rather than a quantitative determination of U, because the mapping assumes U >> t, half filling, and dominant nearest-neighbor superexchange.[89] Deviations from this relation are therefore expected in intermediate or Mott-insulating systems where the relevant excitation manifold is not a simple ligand-to-metal charge-transfer, and materials where systems have different d orbital occupations or mixed valency.[90,91]

**Results and Discussion: Mapping to Phase Diagrams**

Here we consider experimentally constrained and ground-state-derived quantities described within the phase diagrams of the Holstein, Hubbard–Holstein, and t-J–Holstein Hamiltonians. These phase diagrams serve to visualize the influence of different parameters on properties such as polaron strength, carrier localization, on-site repulsions, and spin exchange. We map the measured iron oxide materials to identify different material properties that modulate polaronic parameters, thereby informing design rules. We note that these phase diagrams serve as schematic frameworks to visualize the effects of changes to the parameters from the previous section, rather than quantitative predictions of polaronic behavior. Quantitative quantum dynamics, density functional theory, and other excited state theory is still necessary to definitively ascribe polaronic behavior beyond a first approximation and remains challenging to couple to excited-state experiments. The uncertainties shown in Figures 3-5 reflect propagated experimental uncertainties and do not encompass the systematic uncertainties associated with DFT-derived parameters.

For the Holstein phase diagram, we use two scalar ratios. The first relates to the polaron's strength, $g^2/2dt\omega_0$. The second is $\omega_0/2dt$, which describes the mobility of the polaron via phonons versus free-carrier conduction. Based on Table 1 and the scalar ratios, we construct a phase diagram mapping the measured and derived parameters within the stated approximations in Figure 3. We consider the atomic limit to be the upper right corner of Figure 3, where carriers are completely localized. In the bottom left of Figure 3, photoexcited carriers approach a free carrier limit. We can make a general yet representative cutoff between large and small polarons based on when Equation 4 approaches zero. This is a blurred, approximate transition that is widely debated in the literature, but it is depicted here for simplicity and as a guide to the eye so that the polaron strength can be contextualized in terms of extent of carrier localization. Notably, because the axes of the phase diagram are based upon dimensionless values, the polaron crossover size is not determined explicitly by the value of $E_{polaron}$, but by the balance between lattice trapping and electronic delocalization.[6,92,93]

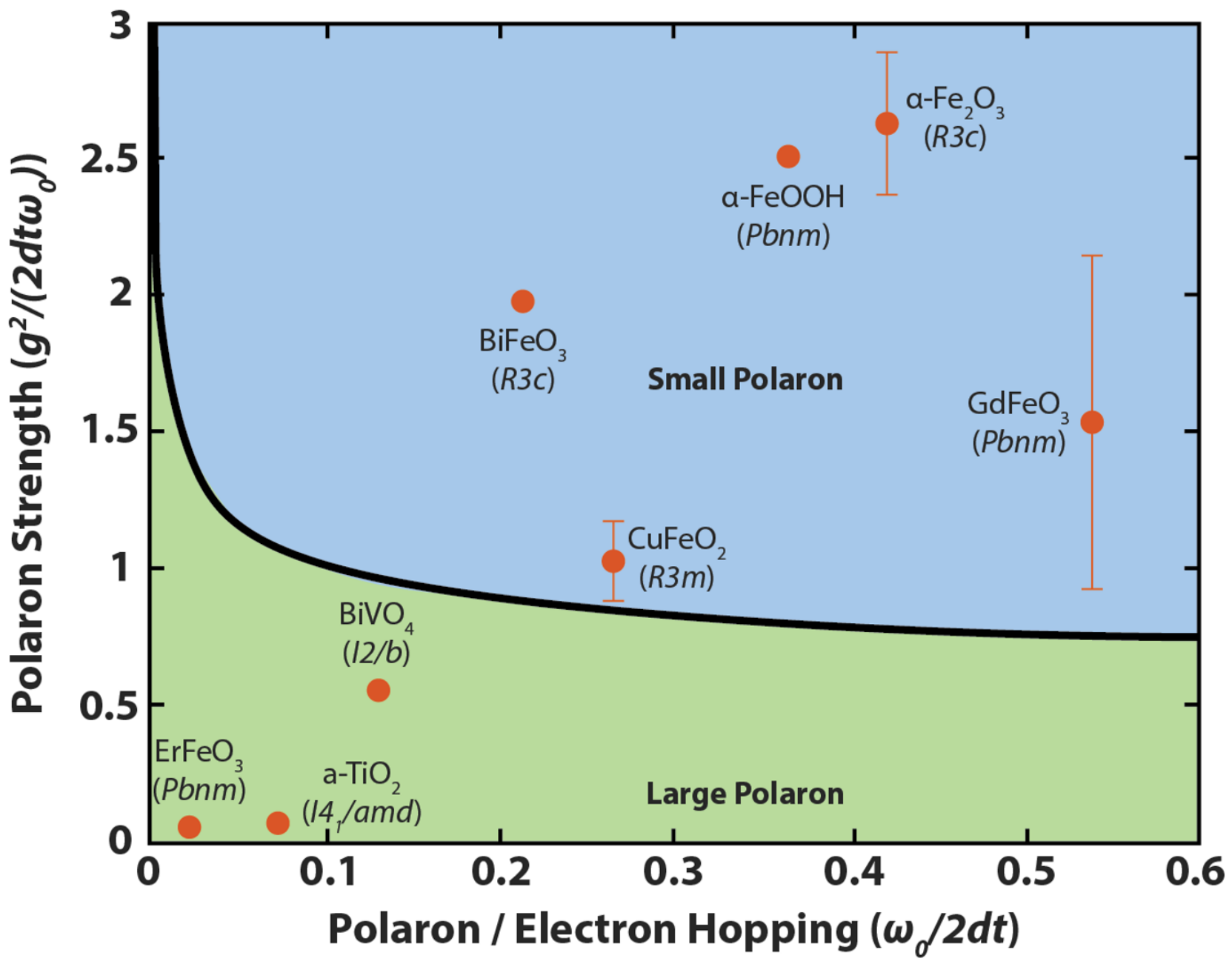

**Figure 3. Mapping measured polarons onto a Holstein Hamiltonian's phase diagram.** Materials and their symmetry group are shown next to the orange dots. The transition between small (blue) and large (green) polarons is schematically denoted by color shading and the solid black line. The polaron size crossover is based on the dimensionless ratios shown on the axes and should not be interpreted as a hard cutoff in the absolute value of $E_{polaron}$. The upper right corner that we attribute to the atomic limit where carriers are strongly localized depicts a regime where $t << g$, while the lower left corner denotes a free conduction regime in which $t >> g$.

The circles in Figure 3 represent transient XUV data from References [17,20,42,44,64,71,94] and optical transient absorption spectra for $BiVO_4$.[82] The agreement between measured polaron properties, where the material falls on the phase diagram with respect to the scaling factors, and the small-to-large polaron crossover is found to be qualitatively predictive. For example, α-$Fe_2O_3$ and α-FeOOH are found to be in the upper right corner near the limit of strong, small polarons and low mobility, which has been widely experimentally demonstrated.[35,95,96] $CuFeO_2$, which has been found to have a polaron with a weaker binding energy than hematite and goethite due to a coherent c-axis lattice expansion that delocalizes the polaron, falls into the middle of the diagram. Spatially resolved measurements and density functional theory which models the extent of carrier localization in $CuFeO_2$ confirm it forms small polarons, in agreement with the assignment in **Figure 3**.[97] $BiVO_4$, which is often debated as having an intermediate-sized polaron, falls on the left boundary close to the crossover between small and large polarons.[98,99] $ErFeO_3$ can be found in the lower left corner because $t \gg g$ due to strong superexchange interactions, which result in a strong electron hopping integral ($t$) and weakened electron–phonon coupling strength ($g$).[100] Spatially resolved measurements of $ErFeO_3$ or ab initio theory determining its polaron size have not yet been undertaken to confirm its size, indicating that further experiment and theory will be necessary to confirm polaronic size in this material.[44] Anatase $TiO_2$, a large polaron material, is also shown for comparison and agrees with the small-to-large polaron crossover depicted in the phase diagram. It is notable that the Holstein model does not well describe materials in the free-carrier limit as $E_{polaron} \sim 2dt$.[7,12]

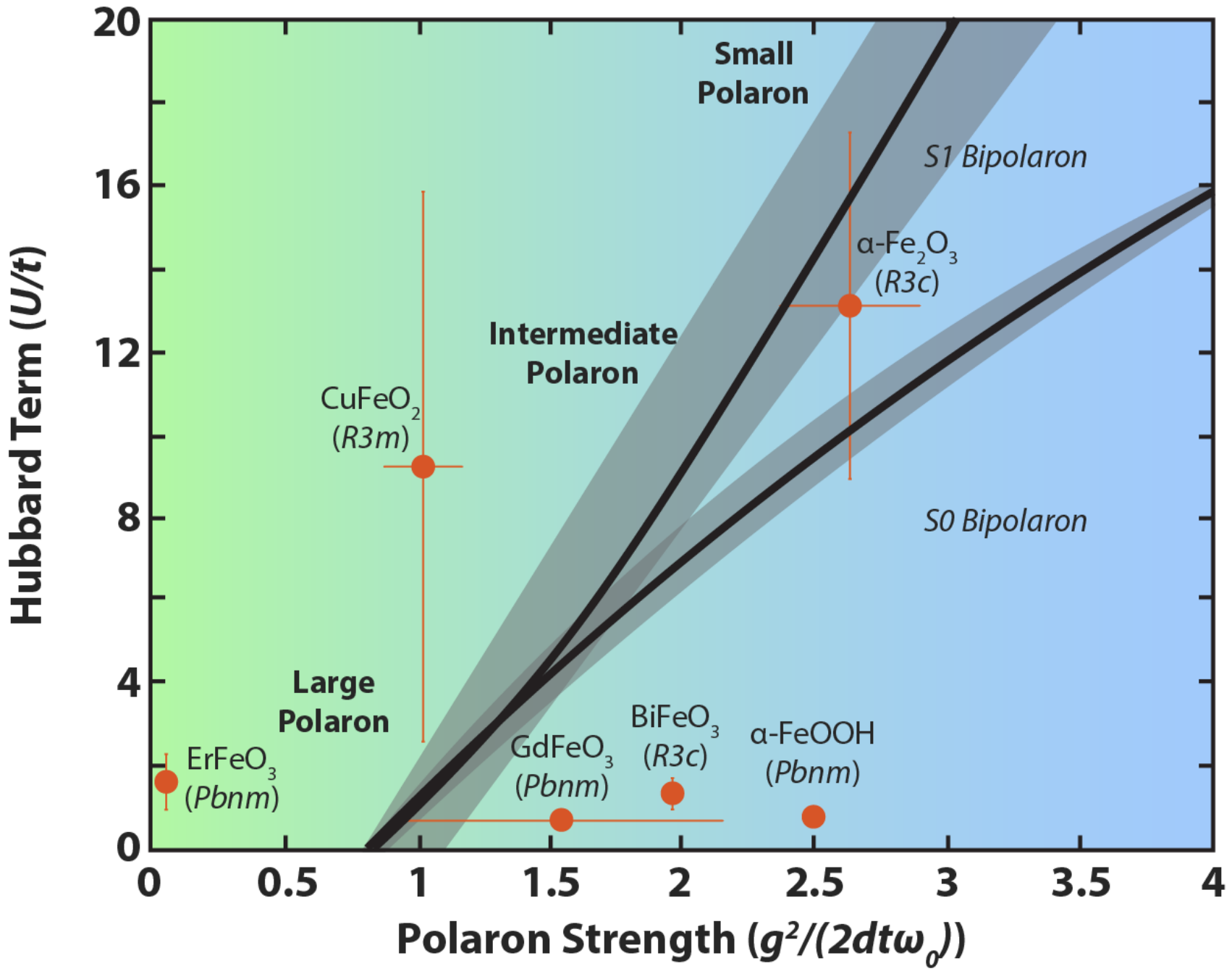

**Figure 4. Mapping experimentally measured polarons to the diagram of the Hubbard–Holstein model.** Small (blue) to large (green) polaron behavior is depicted by the color shading. The black dotted line depicts the boundary between individual polarons and bipolarons. Orange dots depict each measured iron oxide. We do not include the measured materials for which there are no reported J values. The S1/S0 regions indicate where the model predicts bipolaron behavior. The lack of measured bipolarons highlights over-localization that is predicted within the Hubbard–Holstein model. Adapted with permission from reference [101]. Copyright 2004 by the American Physical Society.

We next increase the complexity of our phase diagram by considering a two-particle Hubbard–Holstein model in Figure 4. Only materials with a reported $J$ value, as given in Table 1, are shown. We do not consider literature reported $U$ values as they vary widely depending on the level of theory and functionals applied. The Hubbard parameter acts in multiple regimes. On the phase diagram, it is plotted on the left-hand side by using the relation $J = 4t^2/U$ where $J$ is the exchange energy per spin. We consider the experimentally derived $J$ values to be more reliable reporters of exchange and on-site repulsions, as selecting accurate $U$ parameters in DFT remains a challenge. We note that this is only a first-order approximation, since the relation between $U$ and $J$ depend on many-body wave functions. On the x-axis the polaron strength is parameterized the same as the y-axis of the Holstein model. Depending on the ratio of $U$ to polaron strength, the Hubbard-Holstein model predicts a range of large to small polarons in both extended and bipolaron states. For a relatively small electron–phonon coupling strength, the polarons exist independently of each other. Under the strong electron–phonon coupling limit, the U term is overcome, and bipolarons are predicted either on the same site (S0) or neighboring bound sites (S1).

In the weak-coupling region, where U dominates the electron–phonon coupling strength, we find that the model accurately maps the small-to-intermediate-to-large polaron regime for the single-polaron case, providing more detail than the Holstein model alone. On the other hand, in the strong-coupling region, bipolarons are predicted but are not observed experimentally, resulting in an inaccurate mapping. For example, spatially resolved measurements of hematite down to the quasiparticle limit do not support the presence of S1 bipolarons on neighboring bound sites.[95] This suggests that as polaron strength increases with respect to $U$, an overestimation of the carrier effective mass suggests bipolarons when they are not observed experimentally. Notably, when the Hubbard model is applied to superconductivity and two-dimensional materials, it is found to overestimate the carrier's effective mass in the strongly electron–phonon-coupled limit, due to over-localization.[102–104] This supports the view that the Hubbard–Holstein model overestimates bipolarons and, while a sufficient framework for systems with weaker electron–phonon coupling, may not be sufficient for describing those at the strong electron–phonon coupling limit.[105–107]

However, including the Hubbard term for these transition-metal oxides yields a finer qualitative delineation of the single-polaron species on the left side of Figure 4 and their classification.

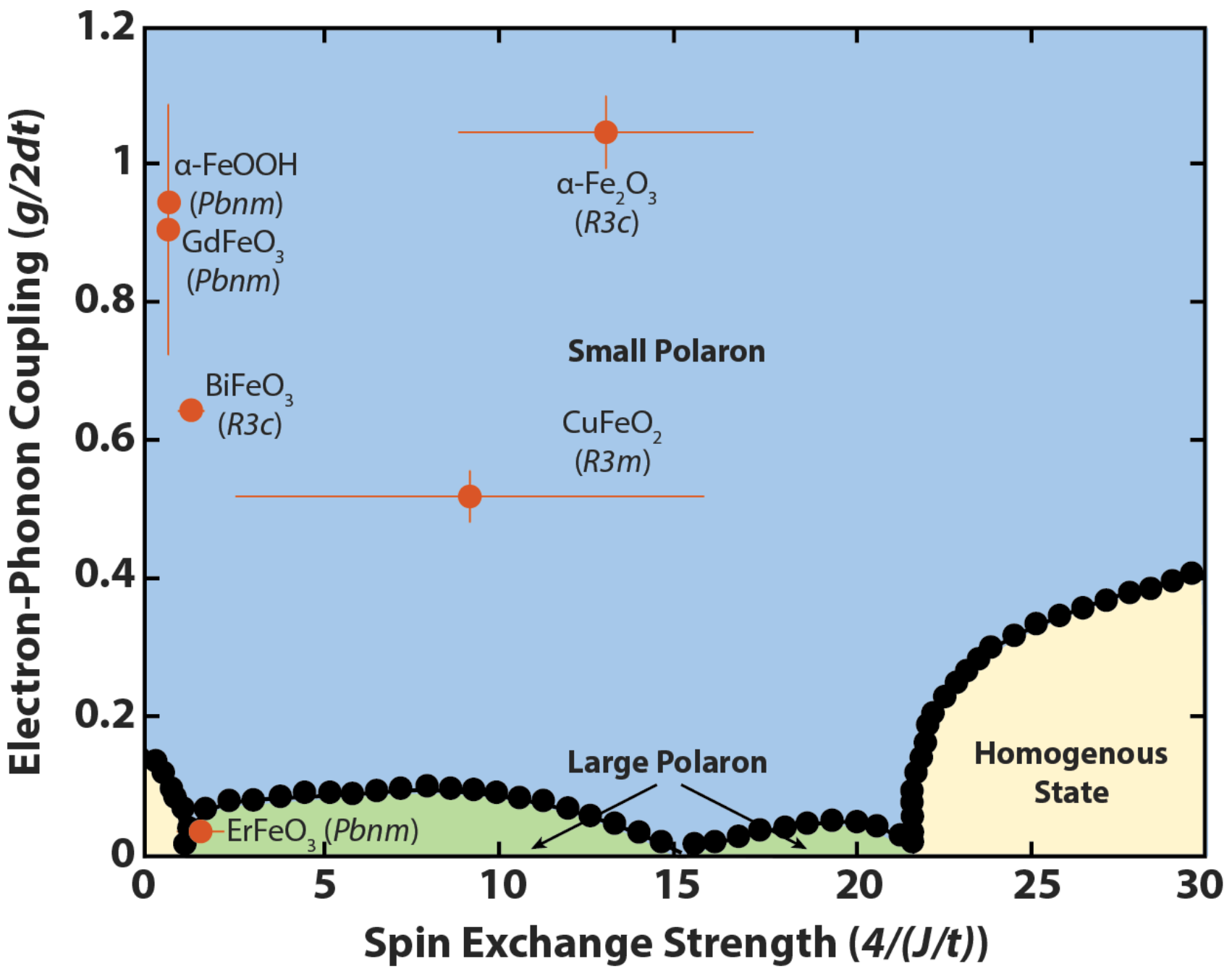

**Figure 5. Experimentally derived polaronic parameters mapped to a phase diagram of the t-J–Holstein model.** Regions corresponding to small polaron (blue), large polaron (green), and near-free carrier (yellow – homogeneous state) behavior are shown by shading. Materials are represented by an orange dot and labeled to include their symmetry group. Adapted with permission from reference [108]. 

We then map the measured iron oxides onto the phase diagram of the t-J–Holstein Hamiltonian for the materials where measured $J$ values exist in literature (Table 1). Figure 5 shows the phase diagram for the ratios of $g/2dt$ and $4/(J/t)$. The parameters $g$ and $J$ must be scaled by that material's $t$ for comparison to the phase diagram of Reference [108]. Figure 5 compares how polaron formation varies as $J$ is modulated for the measured materials. Areas corresponding to small polaron, large polaron, and near-free carrier conduction (homogeneous state) are mapped. Notably, the homogeneous state is not measured in any of the materials presented here, but may be observed in a less polar material. Measurements of a larger materials library are necessary to observe polaronic properties in each regime of these phase diagrams.

The materials that form small polarons mostly agree with the small polaron designation on the phase diagram, while $ErFeO_3$, with its balance of charge hopping and spin exchange, falls into the large polaron regime of the phase diagram. While measurements of $ErFeO_3$ have suggested small polaron behavior, spatially resolved measurements or ab initio theory determining the polaron size must be undertaken to confirm its size.[44] Given that $ErFeO_3$ has polaron binding energies an order of magnitude weaker than hematite and on par with a-$TiO_2$, which is known to form large polarons, further studies are merited to understand the size of polarons within this material.

The diagram depicts that tuning the spin exchange interaction is an effective, though not necessarily straightforward, way to control the small polaron. It must be balanced against the strength of the electron–phonon coupling parameter. Considering these two parameters together reveals that tuning the spin exchange can lead to a rich range of polaron sizes, dynamics, and strengths, extending beyond the electron–phonon parameter of the Holstein model. Superexchange depends on both the Fe–O bond length and the Fe–O–Fe bond angle, as well as spin interactions, providing general guidelines for polaron-based material design. This schematic depiction achieved by mapping experimentally derived parameters to ground-state-theory-based phase diagrams suggests that exchange is a promising route by which polaronic properties may be modulated.

**Conclusions**

We find that the Holstein, Hubbard–Holstein, and t-J–Holstein models provide useful qualitative frameworks to classify excited state electron polaron dynamics probed through metal-centered XUV spectral signatures across a range of transition metal oxides. We find that experimental transient XUV experiments can be mapped onto the parameter space defined by each Hamiltonian's framework and schematically applied to phase-space diagrams previously limited to theory, yielding qualitative agreement between experiment and theory. In particular, the t-J–Holstein Hamiltonian yields the clearest set of design parameters by incorporating spin degrees of freedom and providing the most accurate qualitative distinctions between small and large polarons in the measured iron oxides. Although this is of particular interest for iron oxides with half-filled d orbitals, the design parameter can be generalized to other transition-metal oxides for which measured superexchange interactions are available.[109,110] We demonstrate that tuning electron–phonon coupling strength (*g*), electron localization (*t*), and spin exchange (*J*) provide parameters for understanding and controlling electron polaron formation and localization in the transition-metal oxides examined here. Because these values are interrelated, one must balance them to optimize control over polarons. This work combines experimental data with ground-state theoretical models to provide qualitative phase diagrams that inform electron polaron design at metal centers in transition metal oxides and identifies material parameters that can be tuned for polaron modulation, while also highlighting a need for future nonequilibrium theory and ultrafast experiments to fully connect ultrafast dynamics to ground-state polaronic predictions.

ASSOCIATED CONTENT

The Supporting Information is available free of charge.

- Uncertainty propagation across values in Table 1, singular value decomposition and kinetic fitting model of XUV spectra.

AUTHOR INFORMATION

**Corresponding Author**

Scott K. Cushing - Division of Chemistry and Chemical Engineering, California Institute of Technology, Pasadena, California 91125, United States; Email: scushing@caltech.edu

## ACKNOWLEDGEMENTS

We thank Professor Andrés Montoya-Castillo for helpful discussions and feedback during the preparation of this manuscript. This material is based on work performed by the Liquid Sunlight Alliance, which is supported by the U.S. Department of Energy, Office of Science, Office of Basic Energy Sciences, Fuels from Sunlight Hub under Award Number DE-SC0021266. This research used resources of the National Energy Research Scientific Computing Center, a DOE Office of Science User Facility supported by the Office of Science of the U.S. Department of Energy under contract no. DE-AC02-05CH11231 using NERSC award BES-ERCAP0024109. The computations presented here were, in part, conducted in the Resnick High Performance Computing Center, a facility supported by the Resnick Sustainability Institute at the California Institute of Technology. J.L.M. acknowledges support by the National Science Foundation Graduate Research Fellowship Program under Grant No. 1745301. Any opinions, findings, and conclusions or recommendations expressed in this material are those of the author(s) and do not necessarily reflect the views of the National Science Foundation.

## TOC Graphic

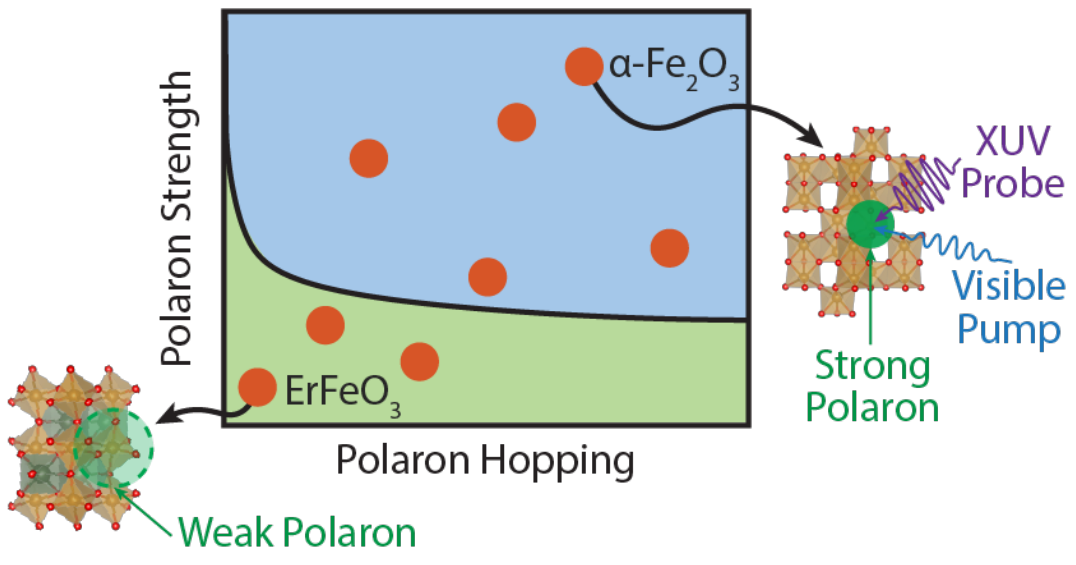

Polaron Strength
Polaron Hopping
α-$Fe_2O_3$
$ErFeO_3$
XUV Probe
Visible Pump
Strong Polaron
Weak Polaron

**Supporting Information for:**

# Experimentally Mapping the Phase Diagrams of Photoexcited Small Polarons

Jocelyn L. Mendes and Scott K. Cushing*

Division of Chemistry and Chemical Engineering, California Institute of Technology, Pasadena, California 91125, United States

*Correspondence and requests for materials should be addressed to S.K.C. (email: scushing@caltech.edu)

# Supporting Information Contents

**S1. Uncertainties in the Polaron Parameterization from Experimental Measurements**

To understand the uncertainties associated with approximating the phase diagrams from experimental observables, we include reported experimental uncertainties for the polaron energy. These experimental uncertainties are propagated through the approximation for $E_{polaron}$ as half of the experimental spectral shift. Several data points from literature do not have reported uncertainties and cannot be statistically analyzed.

We further include reported experimental uncertainties for the experimentally determined exchange energy per spin. In the case where error bars are reported in literature from inelastic neutron scattering experiments, we include those error bars. However, several of the iron oxides studied did not have reported error bars. For these materials, we consider both reported inelastic neutron scattering (INS) and Raman scattering experiments to develop statistics. We also consider theoretically modeled exchange energy where necessary. **Table S1** shows these statistics for several of the material systems under investigation. We note that α-FeOOH has only one experimentally determined value for $J$ in the literature. a-$TiO_2$ and $BiVO_4$ both have $3d^0$ ground state electronic configurations and therefore do not exhibit spin exchange on their transition metal sites.

**Table S1.** Exchange energies from inelastic neutron and Raman scattering experiments and theory. All values are in units of eV. $J$ is approximated as exchange energy per spin. "–" denotes no known literature value for one of the necessary parameters.

| | $J$ (INS) | $J$ (Raman) | $J$ (theory) | Average $J$ value | Standard deviation |
|---|---|---|---|---|---|
| α-$Fe_2O_3$ | 0.0097 ± 0.0031 [1,2] | – | – | 0.0097 [1,2] | 0.0031 [1,2] |
| $CuFeO_2$ | 0.017,[3] 0.014,[4] 0.043[5] | – | – | 0.025 | 0.018 |
| $ErFeO_3$ | 0.38 ± 0.16 [6] | – | – | 0.38 [6] | 0.16 [6] |
| $BiFeO_3$ | 0.16,[7] 0.24,[8] 0.14,[9] 0.12 [10] | – | – | 0.17 | 0.05 |
| α-FeOOH | – | 0.19 [11] | – | 0.19 | – |
| $GdFeO_3$ | 0.10 [12] | – | 0.16 [13] | 0.13 | 0.04 |

Density functional theory (DFT) is known to have systematic uncertainties associated with calculating energies for transition metals. For example, DFT is notorious for underestimating the band gaps of transition-metal oxides in both the local density approximation (LDA) and the generalized gradient approximation (GGA).[14–16] Several methods have emerged to calculate band gaps which agree with experimentally measured values, including the use of an on-site repulsion term (Hubbard U) or the use of hybrid pseudopotentials.[17] Generally though, density functional theory experiences systematic differences depending on the functionals and methods being applied.

All band structures were approximated using the generalized gradient approximation (GGA). On the other hand, phonon dispersions obtained from literature were calculated using different exchange-correlation functionals, ranging from the local density approximation (LDA) to Perdew-Burke-Ernzerhof functionals within the GGA.

**S2. Propagation of Uncertainties Across Derived Parameters**

The uncertainty associated with parameters derived from multiple DFT-based and experimental variables can be propagated according to **Equation S.1:**

$$\sigma_f^2 = \sqrt{\sum_i \left(\frac{\partial f}{\partial x_i}\sigma_{x_i}\right)^2} \quad \textbf{(S.1)}$$

where in this work only experimental uncertainties are considered, as DFT uncertainties are highly method- and potential-dependent, as discussed in the previous section. Therefore, we only consider uncertainty for $E_{polaron}$ and *J*. The uncertainty of the electron–phonon coupling strength (*g*) can be described by **Equation S.2**:

$$\sigma_g = \left|\frac{\partial g}{\partial E_{\text{polaron}}}\right| \sigma_{Epolaron} = \frac{1}{2}\sqrt{\frac{\omega_0}{E_{\text{polaron}}}}\, \sigma_{Epolaron} = \frac{g}{2}\frac{\sigma_{Epolaron}}{E_{\text{polaron}}} \quad \textbf{(S.2)}$$

The polaron strength ($\lambda$) can be described by:

$$\sigma_\lambda = \left|\frac{\partial \lambda}{\partial E_{\text{polaron}}}\right| \sigma_{Epolaron} = \frac{\sigma_{Epolaron}}{2dt} = \lambda\frac{\sigma_{Epolaron}}{E_{\text{polaron}}} \quad \textbf{(S.3)}$$

The uncertainty of the exchange parameter (*J*), normalized by the bandwidth can be written as:

$$\sigma_{4/(J/t)} = \left|\frac{\partial(4t/J)}{\partial J}\right| \sigma_J = \frac{4t}{J^2}\sigma_J = \frac{4t}{J}\frac{\sigma_J}{J} \quad \textbf{(S.4)}$$

and relationally, the on-site Coulomb repulsions can be described by:

$$\sigma_{U/t} = \left(\frac{U}{t}\right)\frac{\sigma_J}{J} \quad \textbf{(S.5)}$$

The propagation of these uncertainties from **Equations S.2–5** are reported in **Table 1** of the main text and are plotted as error bars in **Figures 3–5** in the main text.

### S3. Singular Value Decomposition Kinetic Models

Singular value decomposition (SVD) was employed to extract the spectral weights of different contributions to the transient XUV spectra. The singular value decomposition functions by factorizing the two-dimensional matrix of the transient spectra by rotating (U), scaling (S), and again rotating (V) the matrix. The scaling matrix contains the singular values along its diagonal, which we refer to as the number of components. The number of singular values or components in the scaling matrix is selected to minimize the total number of singular values. The resulting singular values can be applied to specific kinetic features within the transient XUV spectra in agreement with DFT+BSE modeling.

The resulting temporal weights of the SVD components were fit with rate equation models, as described previously, to elucidate the kinetics of the singular values.[18,19] Briefly, a generalized sequential kinetic model is applied to the weighted SVD components of the transient XUV spectra. Here, the sequential kinetic model is always initiated by a charge transfer state ($S_{CT}$) that evolves through one or more intermediate states ($S_n$) before the carrier relaxes. Schematically, this can be understood by **Equation S.6**:

$$S_{CT} \overset{k_1}{\rightarrow} S_1 \overset{k_2}{\rightarrow} S_2 \ldots \overset{k_n}{\rightarrow} S_n \quad \textbf{(S.6)}$$

The rate equations that correspond to these states are dependent upon the number of intermediate states, where in the case of α-$Fe_2O_3$ and $GdFeO_3$, there are only charge transfer ($S_{CT}$) and polaron ($S_P$) states. In the case of $CuFeO_2$ there is initially a charge transfer state ($S_{CT}$), followed by an intermediate polaron state ($S_P$), and finally a *c*-axis lattice expansion state ($S_{Lattice}$). In the case of $ErFeO_3$, there is again an initial charge transfer state ($S_{CT}$), an intermediate $Fe^{3+}$–$Fe^{2+}$ charge hopping state ($S_{Hop}$), followed by the polaron state ($S_P$). **Equations S.7–9** encompass the rate equations for states $S_0$ through $S_n$:

$$\frac{dN_0}{dt} = -k_1 N_0(t) \quad \textbf{(S.7)}$$

$$\frac{dN_i}{dt} = k_i N_{i-1}(t) - k_{i+1} N_i(t) \text{ where } 1 \leq i \leq n-1 \quad \textbf{(S.8)}$$

$$\frac{dN_n}{dt} = k_n N_{n-1}(t) \quad \textbf{(S.9)}$$

where $N_i(t)$ is the time-dependent population of state $S_i$. The lines of best fit in **Figure 2** of the main text are numerical solutions of the differential equations in **Equations S.7–9**, convolved with a Gaussian instrument response function (IRF). The resulting convolved populations were scaled by independent amplitudes and offsets to fit the experimentally determined SVD weights.